# TRAFFIC PERFORMANCE ANALYSIS OF MANET ROUTING PROTOCOL

S.Rajeswari[1], Dr.Y.Venkataramani[2]

[1]Associate professor, Dept. of Electronics and Communication Engineering, Saranathan college of Engineering, rajee_ravi@sify.com
[2]Principal, Saranathan college of Engineering, diracads@saranathan.ac.in.

**ABSTRACT**

*The primary objective of this research work is to study and investigate the performance measures of Gossip Routing protocol and Energy Efficient and Reliable Adaptive Gossip routing protocols. We use TCP and CBR based traffic models to analyze the performance of above mentioned protocols based on the parameters of Packet Delivery Ratio, Average End-to-End Delay and Throughput. We will investigate the effect of change in the simulation time and Number of nodes for the MANET routing protocols. For Simulation, we have used ns-2 simulator.*

## 1. INTRODUCTION

A mobile ad-hoc network (MANET) is a collection of many mobile nodes with no infrastructure. To form a network over radio links, the mobile nodes are self-organized. Extending mobility into the self-organized, mobile and wireless domains is the main objective of MANETs where a set of nodes form the network routing infrastructure in an ad-hoc fashion. MANETs are used in those areas where wired network is unavailable and where rapid deployment and dynamic reconfiguration are necessary. These include military battlefields, emergency search, rescue sites, classrooms and conventions, where the participants share information dynamically using their mobile devices.The topology of the ad hoc network depends on the residual power of the nodes and the mobility and so based on the instantaneous location of the mobile nodes, varying with time [11]. The nodes (each node is acting as a router/transmitter/receiver) are free to move around and organize themselves using random mobility model. Ad hoc wireless networks are self-dependent networks. Hence, they offer unique benefits and flexibility for a variety of situations and applications. Because of these features, the Ad hoc networks are more beneficial than wired network or mobile access. Wireless ad hoc networks consists many nodes which can communicate with each other over multiple wireless hops. The intermediate nodes are used to forward the ongoing traffic. Since the forwarding nodes are fully connected and usually mobile, energy conservation is critical to extend the lifetime of a functioning network. It has been recognized in recent work [1], [2], [3], [4], [5] that, by making the mode of a node as standby or sleep - when the node is idle. So, the energy may be saved up to twofold factor, especially when node density is high. The observation of this concept realizes that nodes consume significant power not only when they are sending or receiving packets, but also when they are idle or overhearing ongoing traffic. It has been studied that energy dissipation of an idle node is equivalent to half of an active transmitting node when it is provided with maximum power. Further, when the node density is high in the network, it has been studied that each backbone node has the capacity of forwarding ongoing traffic as a representative of the local neighborhood within its range. Connectivity is still maintained due to the natural routing redundancy when the node density is sufficiently high. These observations provide sufficient incentives to design algorithms to turn off all locally and spatially redundant nodes in order to conserve energy. Existing work [2], [3] following this concept uses periodic message broadcasts in the local neighborhood to drive election algorithms to make local decisions on if a node should participate in the virtual backbone. In GAF (Geographical Adaptive Fidelity) [2], geographic information of each node is assumed to be available via location sources such as GPS, and geographic states of neighboring nodes are periodically exchanged to construct the backbone. In Span [3], periodic HELLO messages are locally broadcast, and are used to drive local algorithms for the nodes to elect or remove themselves from the backbone. The periodic exchange of local broadcast messages has provided the convenience of designing algorithms based on full knowledge of the states of neighboring nodes.

Recent work [4] proposes to wake up nodes that are necessary for traffic forwarding only, and let other nodes go to sleeping mode. While such a scheme reduces the periodic message exchange overhead, it pays a price in terms of new session setup latency, since nodes along the transmission route of the new session are very likely to be in the sleeping mode.

In this paper, we explain about routing protocols in section 2 and in section 3, we brief about power saving mode of 802.11. We present the proposed work in section 4. In section 5, we are given the simulation results of traffic analysis. Section 6 concludes this paper.

## 2. ROUTING PROTOCOLS

Routing in Mobile Ad-hoc Network has been a subject of extensive research over the past several years. A routing protocol is needed because it has to pass several hops (multi-hop) to ensure that a packet reaches the destination [7].

Routing protocol has two significant functions:

i. Selecting the routes for various source-destination pairs and among those choose the best path ( least cost/ minimum distance/ More bandwidth)

ii. Routing each messages through that best path to their correct destination.

The second function is implemented by a variety of protocols with the help of routing tables. Ad-hoc routing protocols can be classified based on different criteria. Based on the routing mechanism used by a given protocol, it may fall under more than one class. Routing protocols for Ad-hoc networking can be classified [6] into four categories viz.

(i) Based on routing information which are updated by anyone of the routing mechanism (proactive or table-driven, reactive or on-demand and hybrid protocols)

(ii) Based on that instantaneous time information (Both Past and Future time information) for routing

(iii) Based on routing topology (Flat Topology, Hierarchical Topology)

(iv) Based on the use of particular resources (Power Aware Routing and Geographical Information Assisted Routing) for assistance.

Each node in these routing protocols in Mobile Ad hoc networks operates on constrained battery power. The power will start decreases with time even though the node is idle. Power management is an important concept which concentrates how to reduce the energy consumed in the wireless interface of battery-operated mobile devices. So Energy Conservation [14] is taken as a prime factor since all wireless devices usually rely on portable power sources such as batteries to provide the necessary power.

## 3. POWER SAVE MODE IN IEEE 802.11

Towards the direction of powering off nodes periodically, the wireless LAN MAC layer specifications in the current IEEE 802.11 standard support a Power Save Mode (PSM) [9] that conserves energy on idle nodes, by powering their wireless interface off for selected periods of time. Such a Power Save mode is applicable to both infrastructure networks (Basic Service Set), and ad hoc networks (Independent Basic Service Set). In the specification (hereafter referred to as 802.11 PSM), each station may be in one of two power modes: awake (when the node is fully powered) and doze (when the node is not able to transmit or receive, and consumes very little power). The local clocks of nodes that are in the Power Save mode are synchronized by periodic beacon messages. Each beacon message marks the beginning of a beacon interval. At this time, a node should suspend all its backoff timers, and after a random backoff period, attempt to send a beacon message if it has not yet received such beacons from other nodes in the local neighborhood. The one who first transmits the beacon message will prevent others to send any further beacons during the corresponding beacon interval. Each beacon interval starts with an Ad Hoc Traffic Indication Message (ATIM) window. Any data packets destined to a dozing node will be buffered at the upstream neighbor until the upcoming ATIM window. During the ATIM window, all nodes in the local neighborhood will have already been clock-synchronized, and will be awake during the same time interval. Buffered packets will be advertised during the ATIM window by special ATIM frames that we refer to as advertisements. Advertisements for unicast data packets need to be acknowledged, while those for broadcast packets do not. A node that has acknowledged previous advertisements should stay awake for one beacon interval in order to receive buffered data packets from the sender. Successfully advertised (and acknowledged, for unicast packets) data packets can be transmitted after the ATIM window. Packets not successfully delivered will be retried during the next beacon interval.

Using such a pure 802.11 PSM approach, all participating nodes in the ad hoc network stay in the Power Save mode. Consider the setup latency of a new route from the point of view of an on-demand ad hoc routing protocol. Since both unicast and broadcast packets need to be buffered and subsequently advertised before actual transmissions, the setup latency for a new route will be inevitably high. It may be trivially derived that such setup latency is on the magnitude of kT, where k is the number of hops on the route, and T is the length of a beacon interval. One may believe that once the connection is established, all nodes on the route may stay awake during the lifetime of the connection. This is not true without modifications to the current IEEE 802.11 PSM specification. In the current protocol, packets buffered at an upstream node are only marked when an acknowledgment from the corresponding downstream node is received in the ATIM window, and only marked packets can be transmitted in the same beacon interval after the ATIM window. Therefore, a packet that arrives between two ATIM windows has to wait until the next ATIM window to be advertised and marked. This way, the average end-to-end latency of all data packets is on the same magnitude as the setup latency of a new connection. Since $K \alpha O(N)^{1/2}$ , where N is the number of nodes in the network, it is not scalable to the network size. Odds seek to improve both the average end-to-end latency and the connection setup latency.

## 4. A NEW PROPOSED ROUTING PROTOCOL

In this paper, based on the gossip-based ad hoc routing, we propose a Reliable and Energy Efficient Gossip Routing Protocol to achieve energy efficiency and reliability in wireless ad hoc networks to overcome the above drawbacks. In this protocol, the nodes can be in active mode with probability 1-p or sleep mode with probability p which is fixed at the initial stage. A node (which wants to communicate) maintains a control variable called B which represents the current number of neighbors at each node which are kept in active state. The rest of the nodes will be in either p or 1-p state. The higher - B is the more power the node uses to send packets and thus the communication is more reliable. When node X needs to broadcast a data packet, X looks up its neighbor list for the distance between itself and its neighbors numbered B. X then calculates the amount of power needed to send the packet to that neighbor. Initially, every node initializes B to one. This means that a node initially broadcasts data packets only to its closest neighbor, thus requiring the least power. After sending data packet, node X waits for a feedback from destination. While receiving packets at the destination, the delivery ratio D is calculated and it will be sent as a feedback to the source. If X hears a feedback D for the data packet below a reliability threshold RT, X increases the value of B there by increasing the probability of active nodes. This assures the increase in delivery ratio and increase the power consumption. When D becomes greater than or equal to RT, the value of B is decreased adaptively to decrease the number of forwarding nodes and there by decrease the probability of active nodes which will reduce the power consumption. This process continues until either X hears a feedback for the packet or the value of B reaches reliability threshold RT, which is determined by the total number of neighbors. The same procedure is repeated for every arrival of the beacon signal. In polling based power management policies, a node polls other nodes periodically (or equivalently makes announcement to other nodes) to decide whether or not it (or the others) should remain active. One example of this type of policies is IEEE 802.11 power saving mode (PSM). In the IEEE 802.11 specification, all nodes in the network are synchronized to wake up periodically in every beacon interval.

The major objective as proposed in this protocol is to achieve energy efficiency by putting some nodes in a sleep mode. The sleep nodes in the path of the packet transmissions are made to be active state by a triggering signal which will consume some extra amount of power. But we can achieve reliability and energy conservation compared to exiting protocols.And in this protocol, once the threshold value is achieved the nodes are again driven to sleep mode which will conserve some extra amount of power compared to the existing power saving schemes.

## 5. Comparison of TCP and CBR Traffic flow

Information regarding individual node behavior can be used for identifying selfishness behavior in the network. The link between any pair of nodes is established by TCP and as well as by CBR Traffic pattern. NS2 is used for simulation study when nodes are deployed in a high-traffic environment, with multiple connections for the above routing protocol.

## 6. SIMULATION RESULTS

NS2 is used to simulate the proposed algorithm. In our simulation, the channel capacity of mobile hosts is set to the same value: 2 Mbps. For the MAC layer protocol the distributed coordination function (DCF) of IEEE 802.11 (for wireless LANs) is used. It has the functionality to notify the network layer about link breakage.

In the simulation, mobile nodes move in a 600 meter x 400 meter region for different simulation time. The number of mobile nodes is kept as 50. We assume each node moves independently with the same average speed. All nodes have the same transmission range of 250 meters. In our simulation, the speed is set as 20m/s. The simulated traffic is Constant Bit Rate (CBR). The pause time of the mobile node is kept as 10 sec.

## 6.1 PERFORMANCE METRICS

Mobile ad hoc networks have distinguished characteristics because of dynamic topology, time-varying and bandwidth constrained wireless channels, multi-hop routing, and distributed control and management. Traffic performance analysis of this routing protocol for mobile ad hoc network (MANET) is studied with the above stated parameters and the graphs are plotted. This is used to analyze the traffic pattern performance of this routing protocol, which leads to find its suitability and functionality. Specifically, this paper evaluates the traffic pattern performance comparison of GSP (exist) and this proposed new Routing protocols on the following performance metrics: Average end-to-end delay, Packet delivery ratio and throughput with the increasing number of nodes and for varying simulation time.

## 6.2 BASED ON SIMULATION TIME

Performance of AEERG and GSP protocols is evaluated under both CBR and TCP traffic pattern for different simulation time. Extensive Simulation is done by using NS-2.

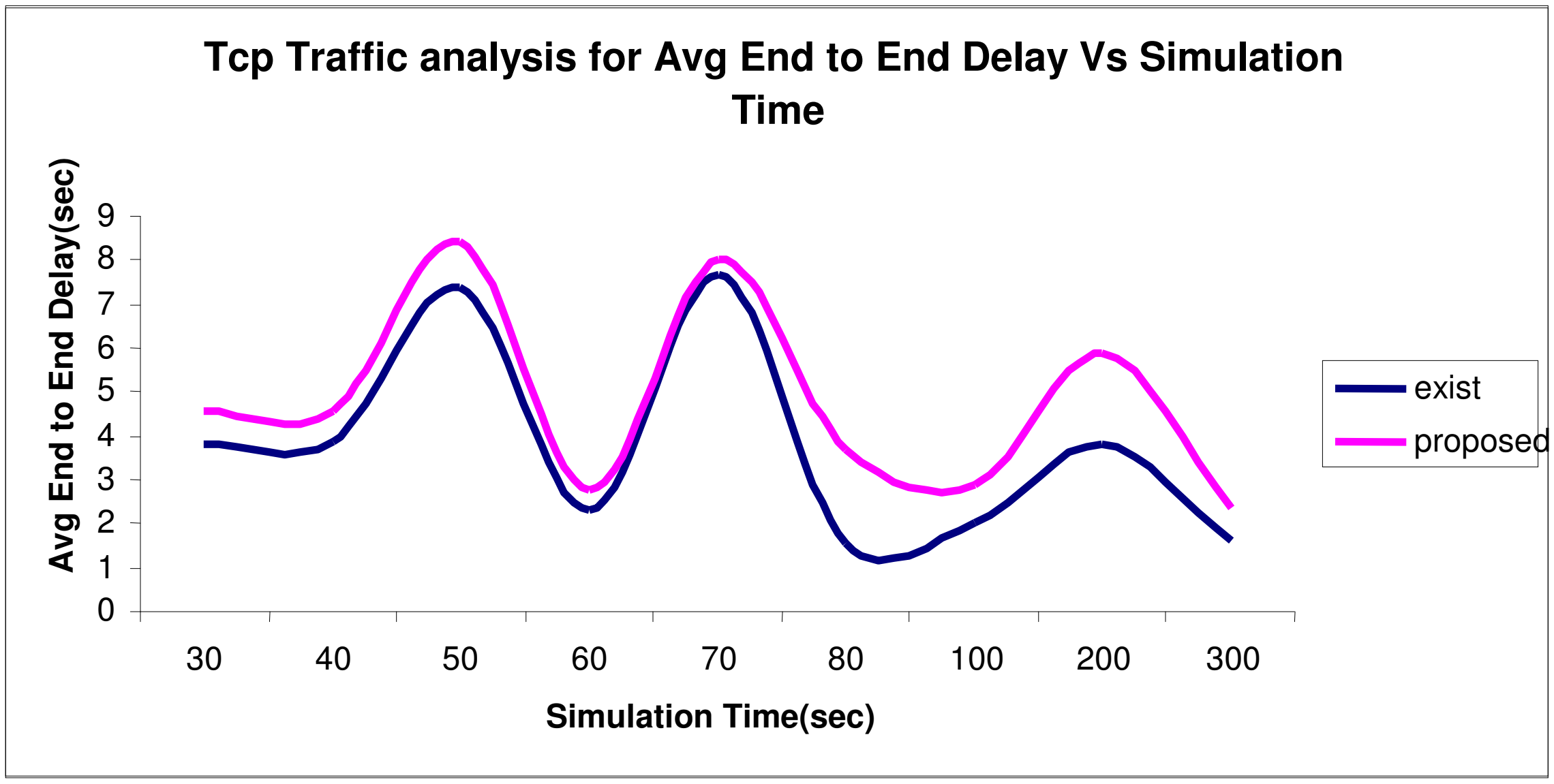

Figure 1 Average End to End Delay TCP Traffic Pattern

Fig I give the Average End to End Delay of protocols for different simulation time for the TCP traffic pattern and the Fig II for the CBR traffic pattern. We can see from the figure, the Average End to End Delay for TCP is more in the case of proposed than exist whereas for the CBR, Average End to End Delay is less in the case of proposed than exist. TCP assures the guaranteed packet delivery, the delay is more. Even though CBR is unreliable, the delay is less.

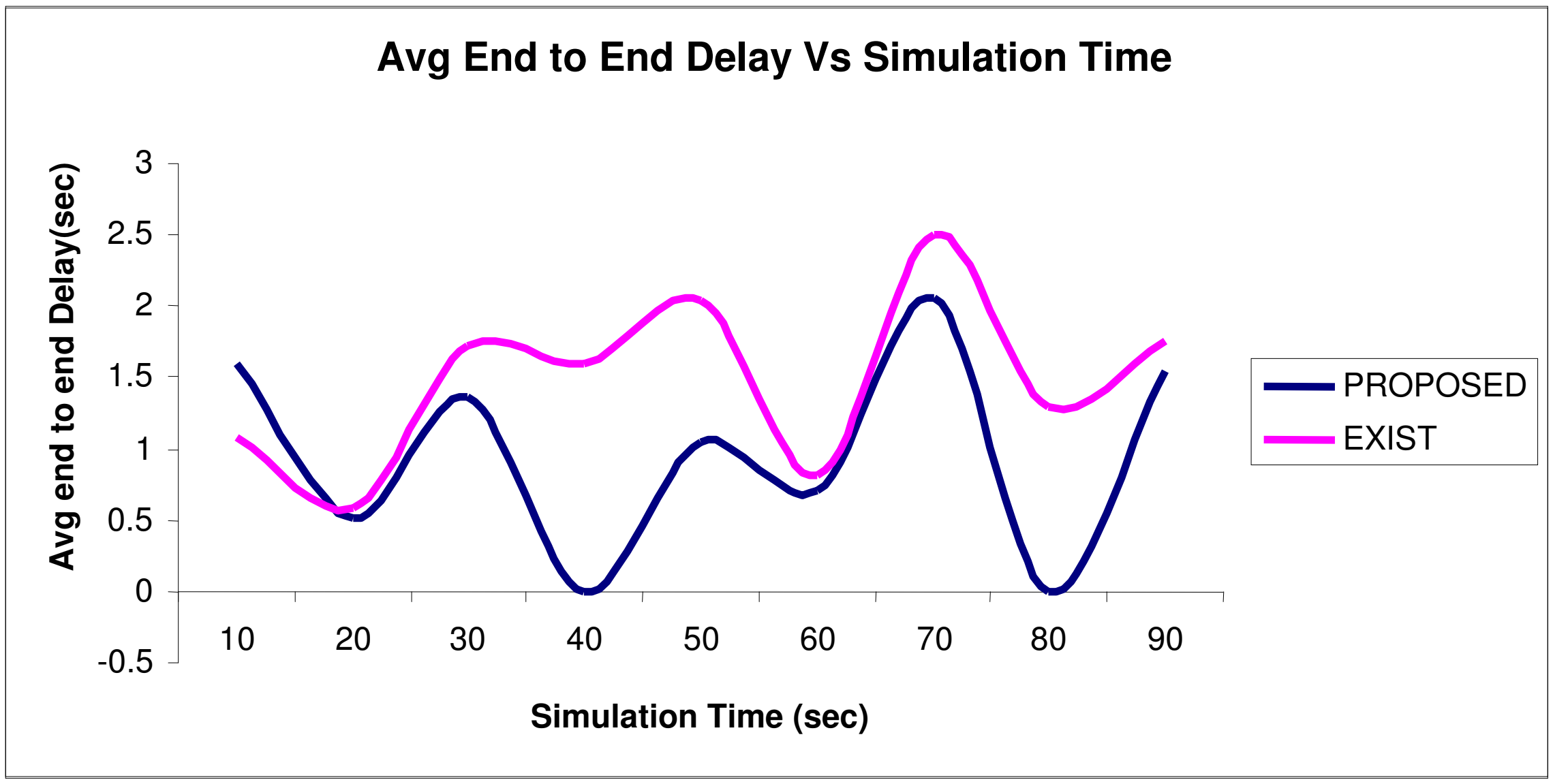

Figure 2 Average End to End Delay CBR Traffic Pattern

Packet delivery ratio is calculated by dividing the number of packets received by the destination to the number of packets sent by the source. It specifies the packet loss rate, which limits the maximum throughput of the network. The better the delivery ratio, the more complete and correct is the routing protocol. If we employ TCP traffic pattern due to the guaranteed packet delivery, the Delivery Ratio is more in the case of TCP than in the case of CBR. But by means of both the traffic pattern, we can understand that the proposed protocol performance is greater than the existing scheme. Fig III and Fig IV give the TCP traffic analysis and CBR traffic analysis respectively for Delivery Ratio with varying simulation time. Fig V and Fig VI gives the throughput for different simulation time for TCP and CBR respectively.

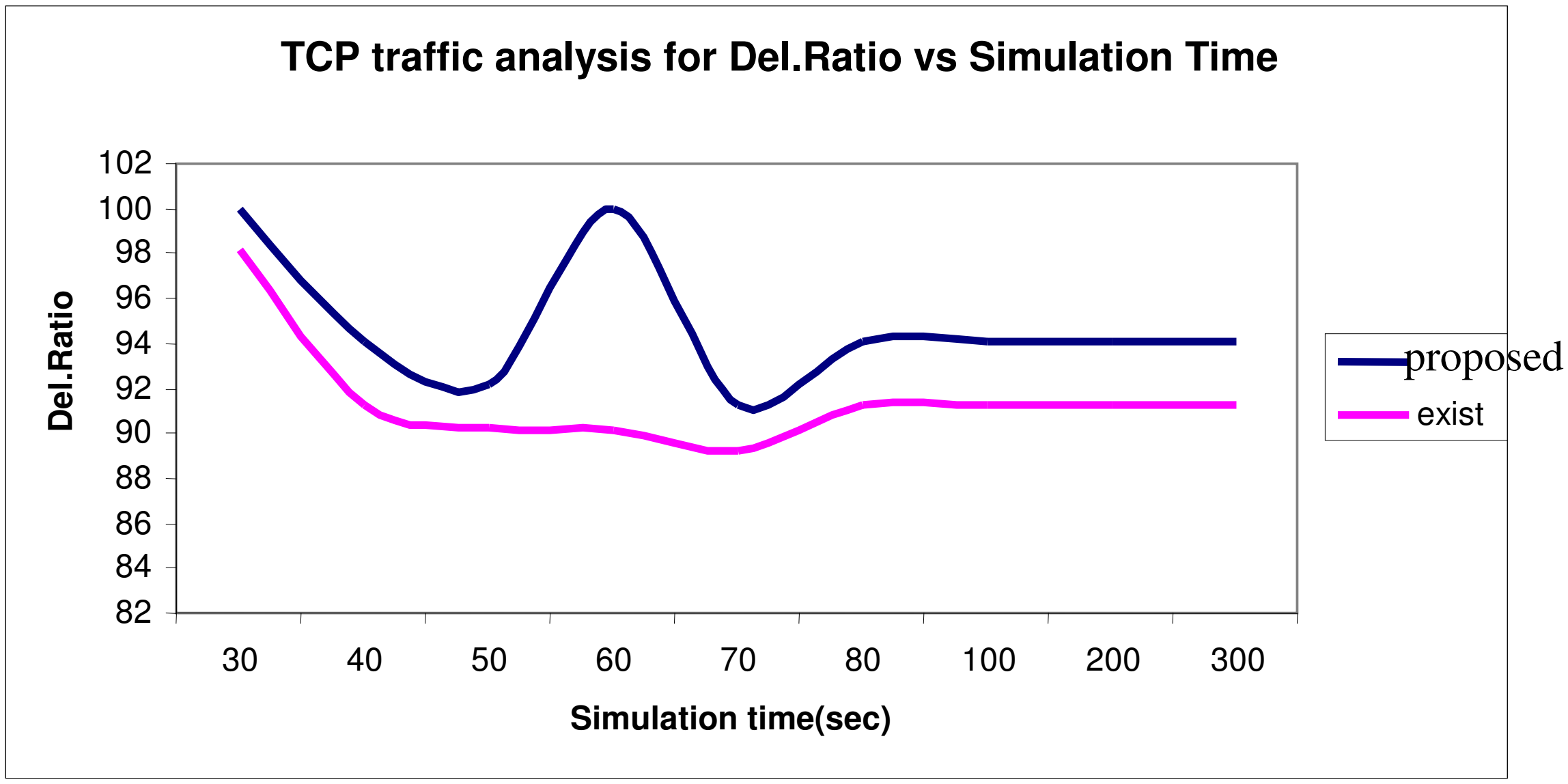

Figure III Delivery Ratio TCP Traffic Pattern

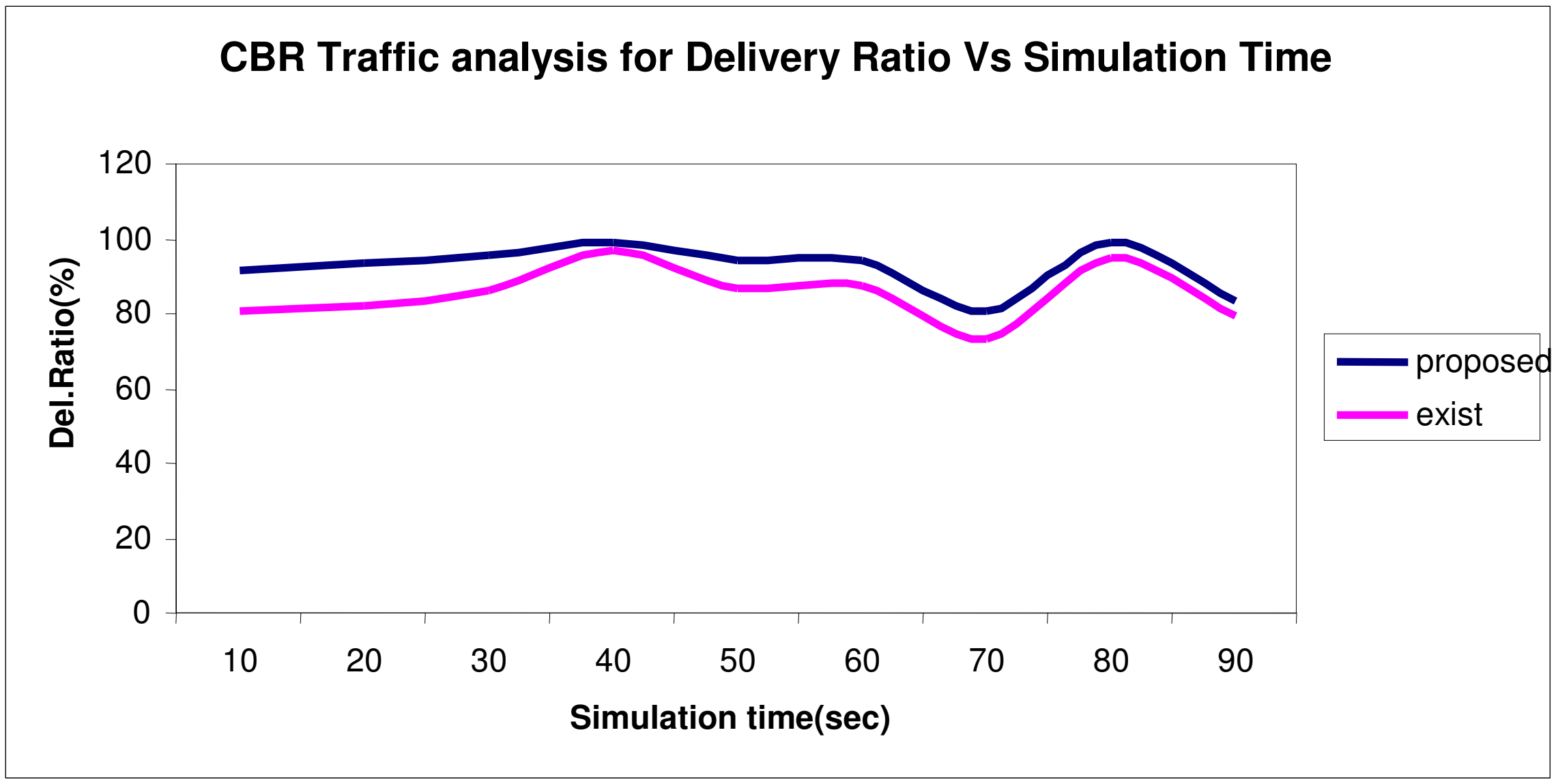

Figure IV Delivery Ratio CBR Traffic Pattern

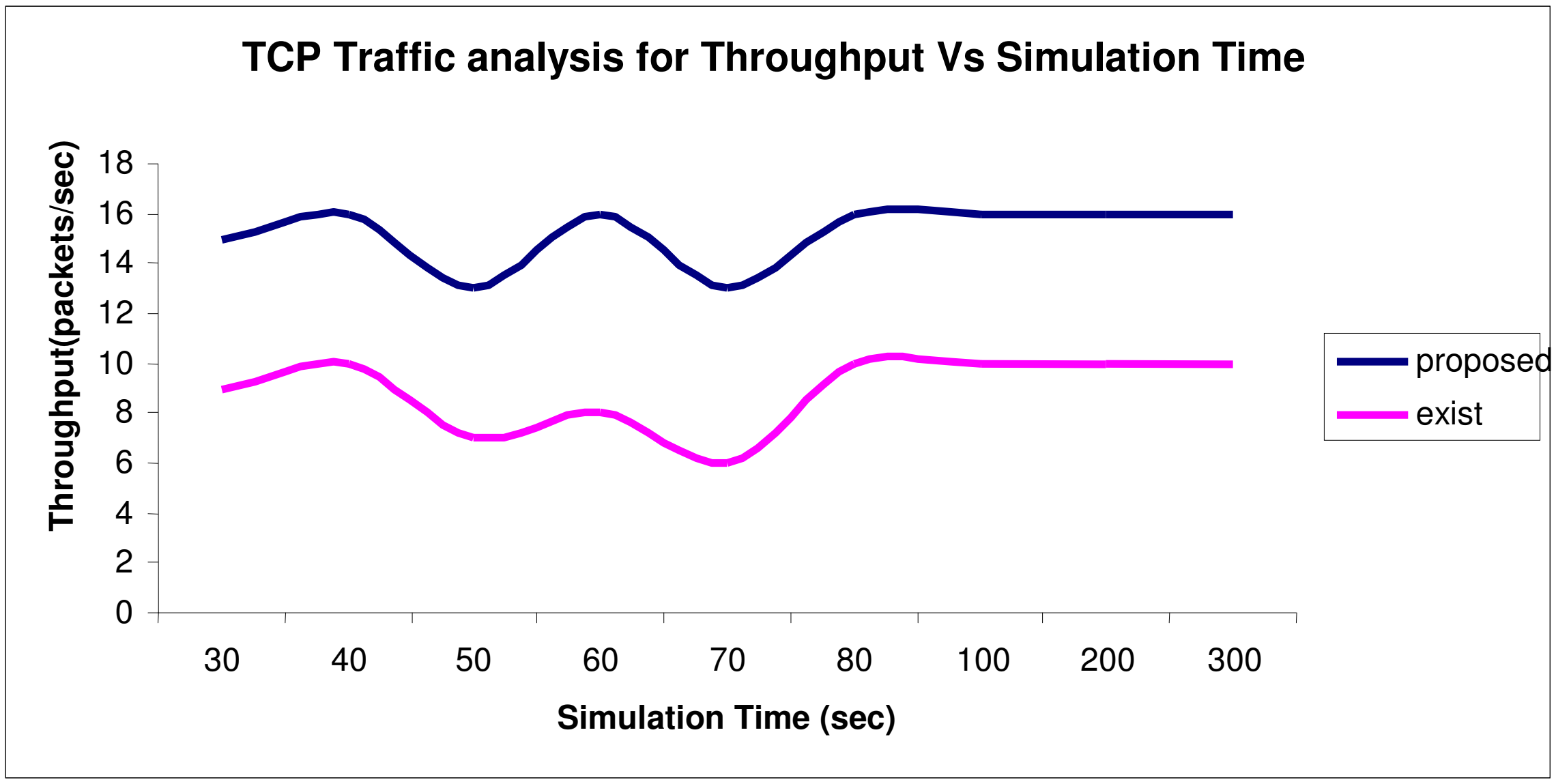

Figure V Throughput TCP Traffic Pattern

Fig V gives the Throughput of protocols with varying the simulation time for TCP and Fig VI give Throughput for CBR traffic pattern. We realized that the Throughput of proposed scheme is more than the existing scheme. For simulation time less than 100 sec, the successful number of packet received is varied, whereas for above 100 sec of simulation time, the throughput is constant for both the schemes.

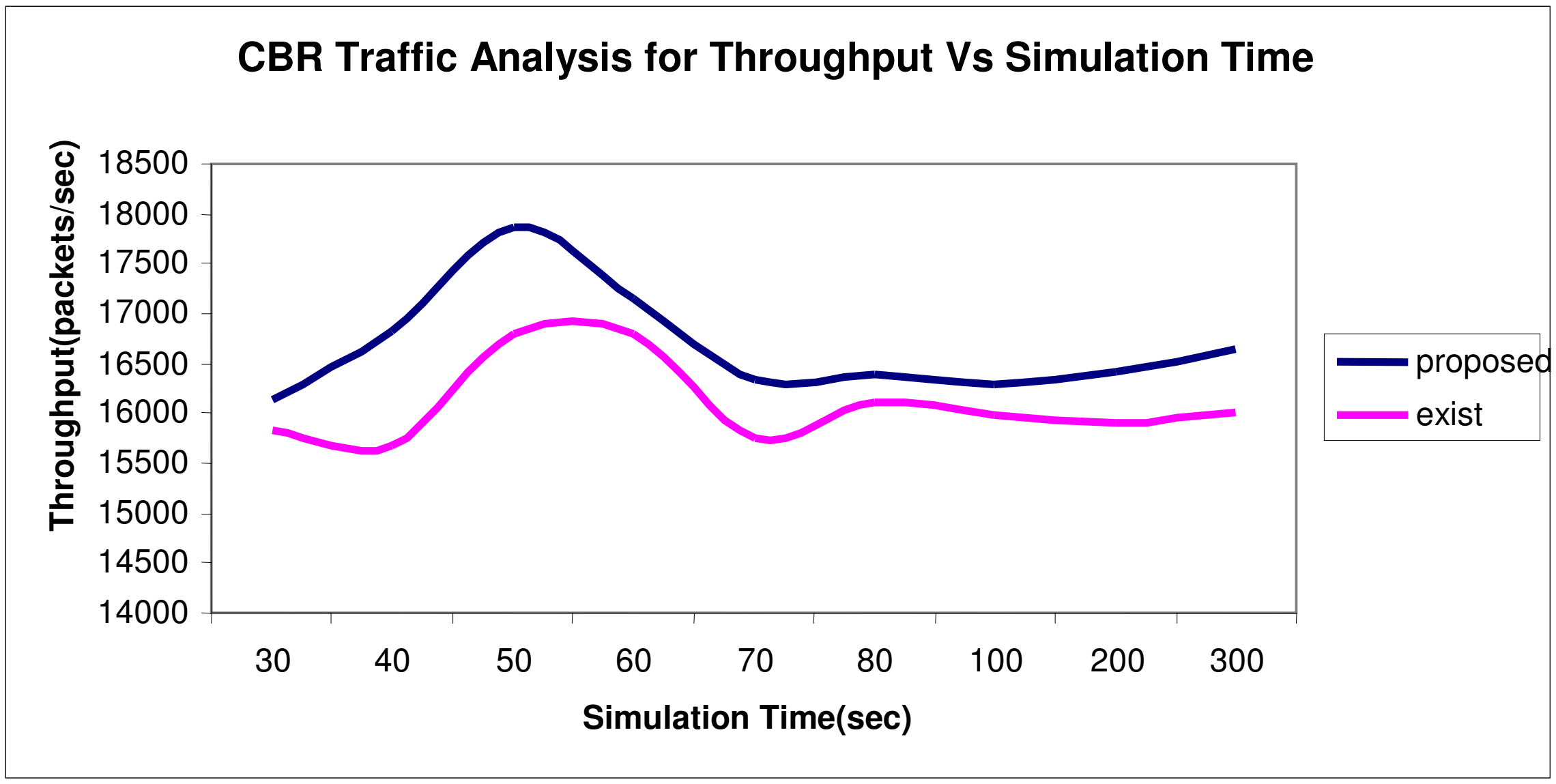

Figure VI Throughput CBR Traffic Pattern

## 6.3 BASED ON VARYING NUMBER OF NODES

Performance of proposed and existing protocols is evaluated under both CBR and TCP traffic pattern for varying number of nodes. Extensive Simulation is done by using NS-2.

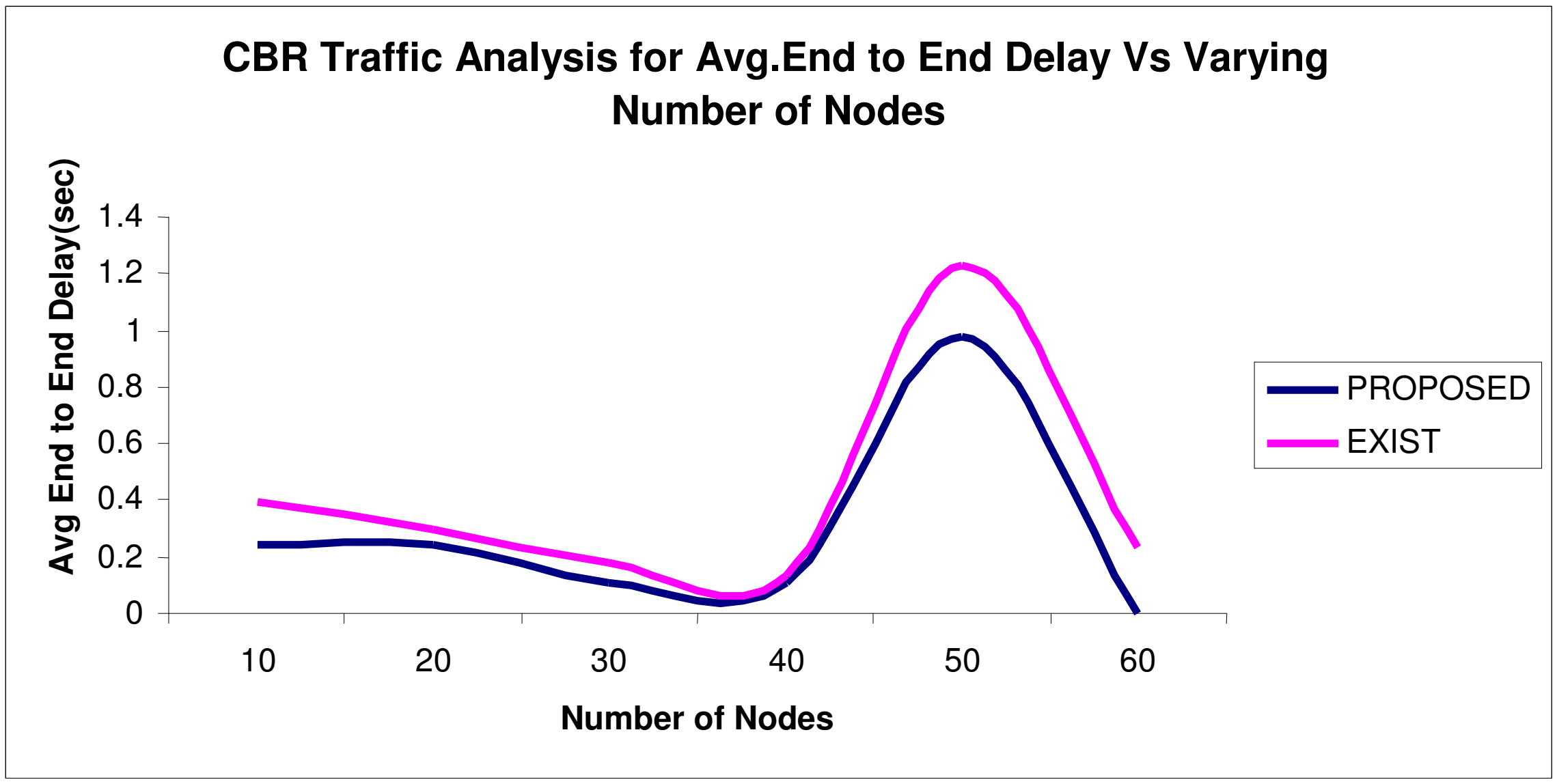

Figure VII Average End to End Delay CBR Traffic Pattern

The same experiment is repeated with increasing number of nodes. Fig VII gives the Average End to End Delay of protocols when the Number of node is increased. As we observe from the figure VII and VIII, the Average End to End Delay of is less in both the traffic patterns when it is compared with the existing scheme. But due to increasing number of nodes, the delay is also increased in both the cases up to 50 numbers of nodes. Since the adaptive nature of the protocol, beyond 50 numbers of nodes the delay starts decreasing.

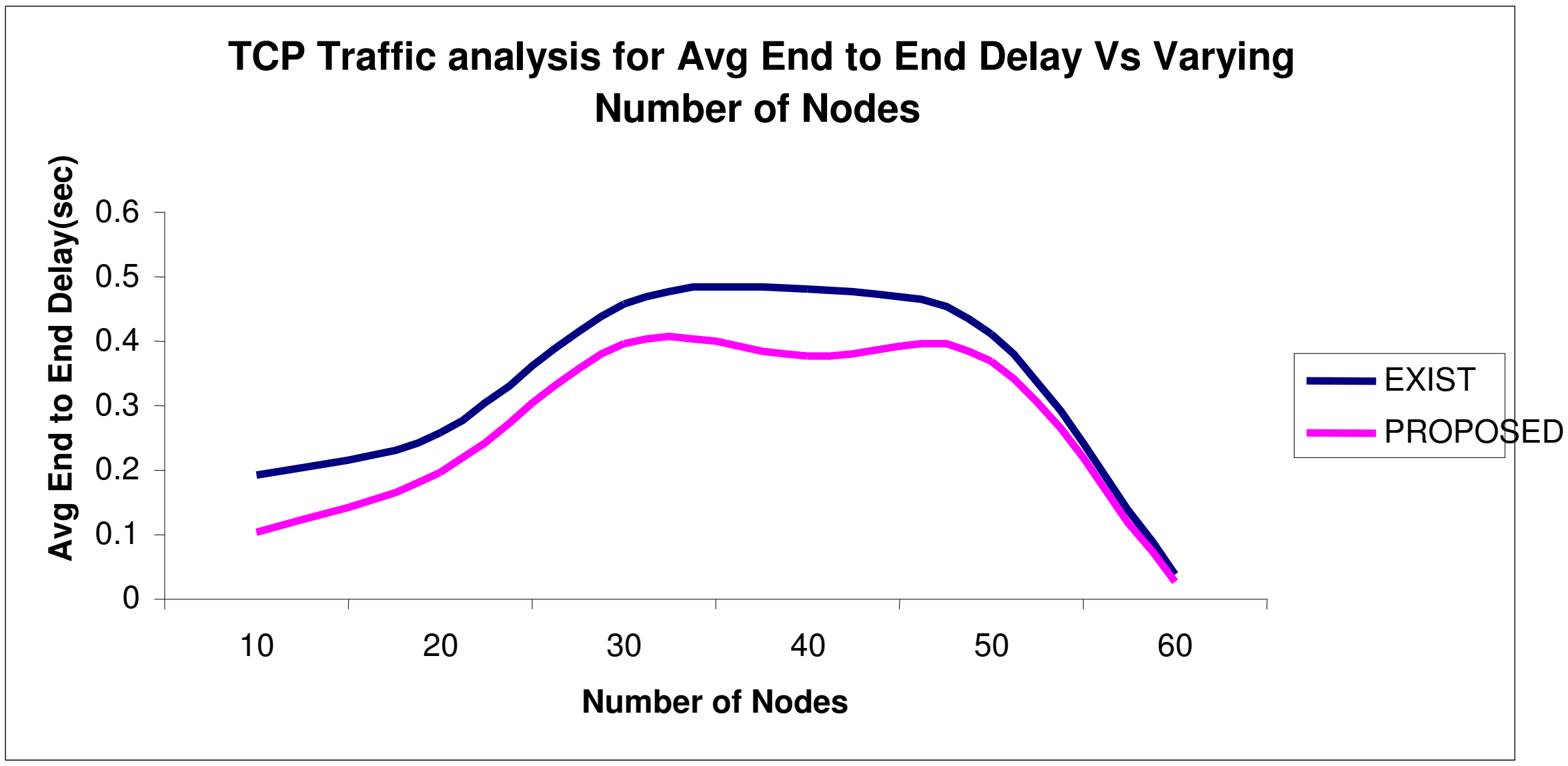

Figure VIII Average End to End Delay TCP Traffic Pattern

From Figures IX and X, we can understand the throughput is increased for both the traffic pattern compared to the existing scheme. Above 50 numbers of nodes, the throughput will start increasing. In CBR, after it reaches one point due to congestion the throughput will start decreasing whereas in the proposed scheme due to the adaptive nature the throughput is increasing linearly. The same is shown for the results of Packet Delivery Ratio for CBR and TCP traffic pattern in XI and XII figures.

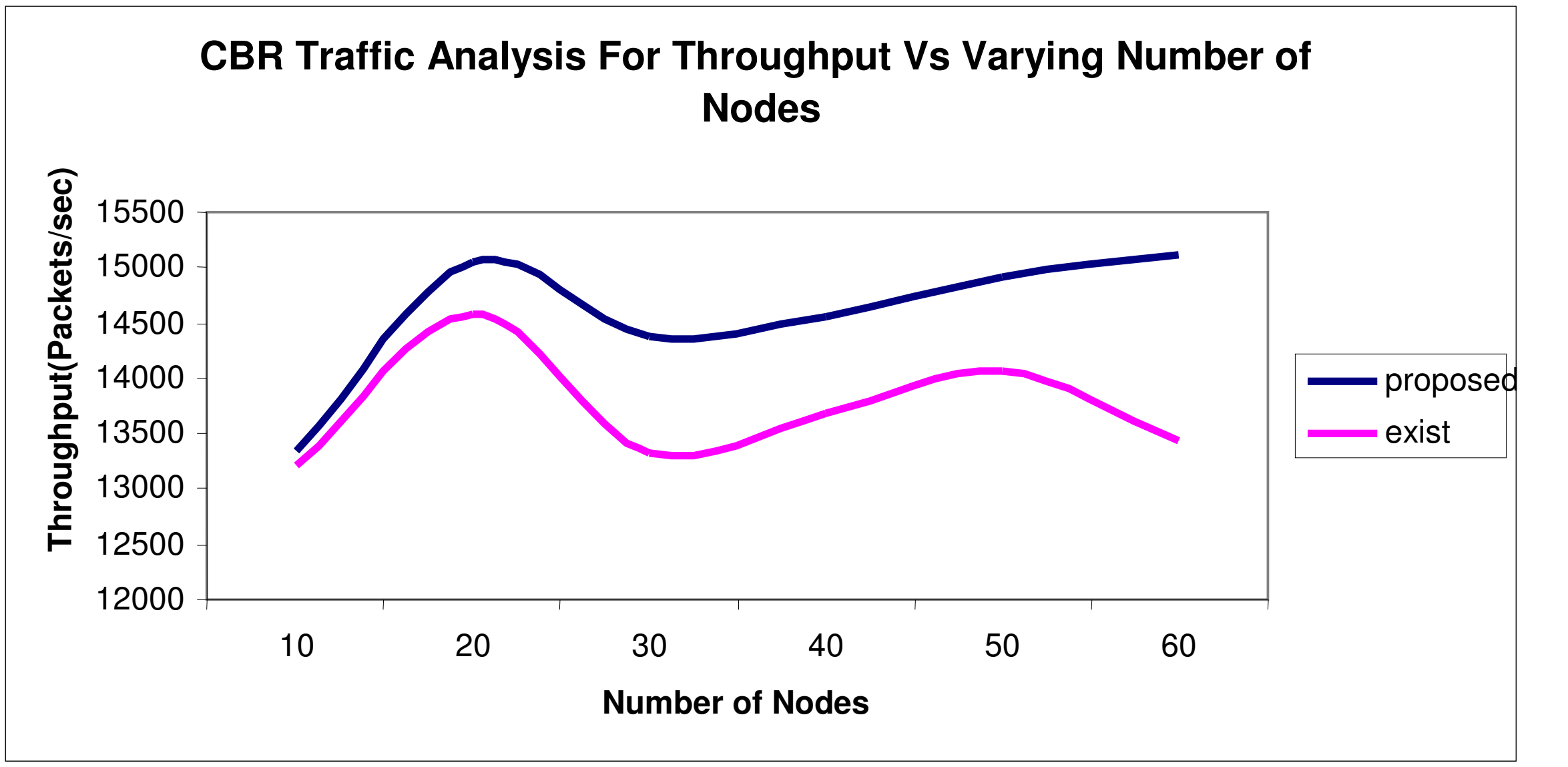

Figure IX Throughput CBR Traffic Pattern

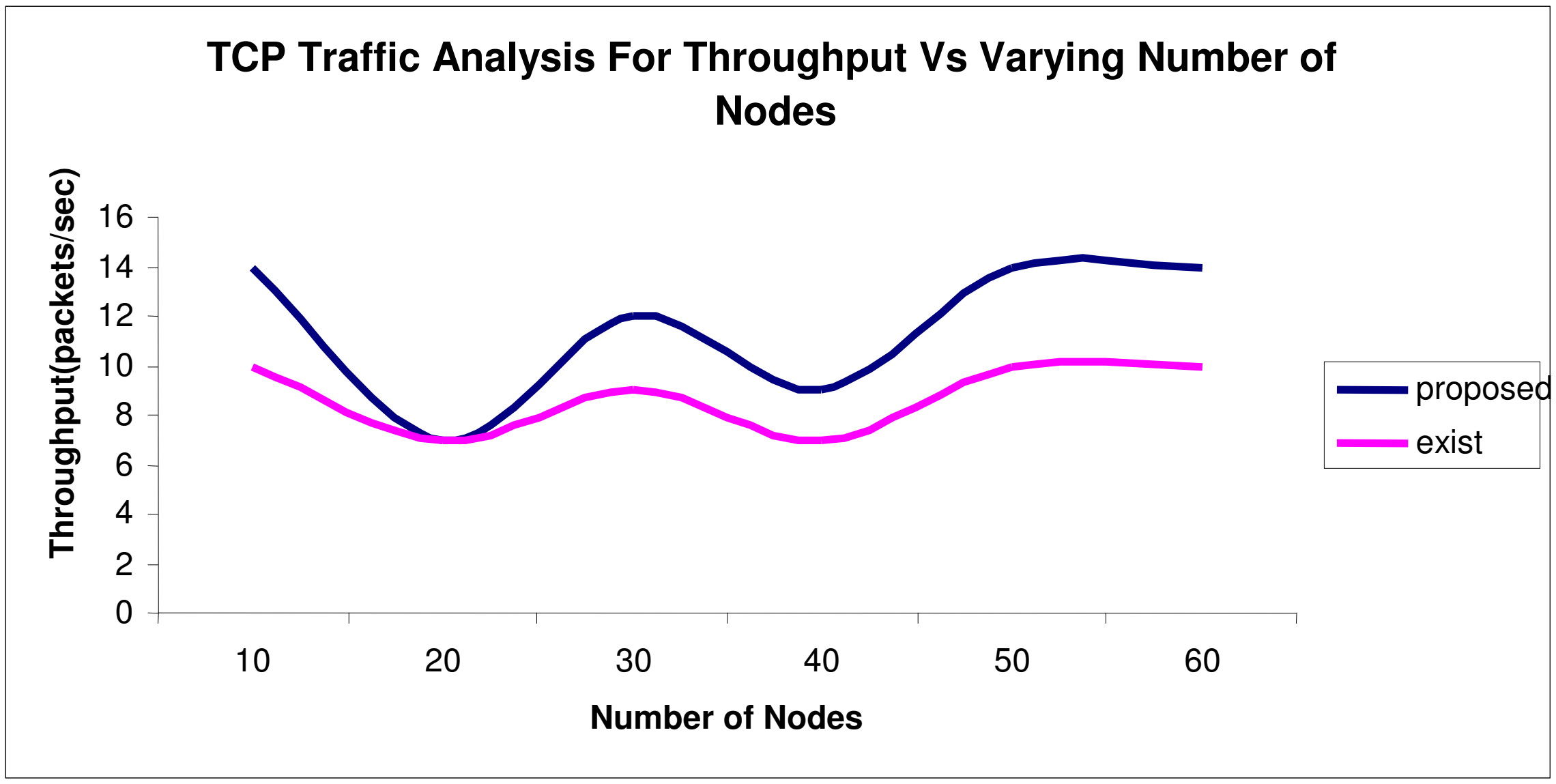

Figure X Throughput TCP Traffic Pattern

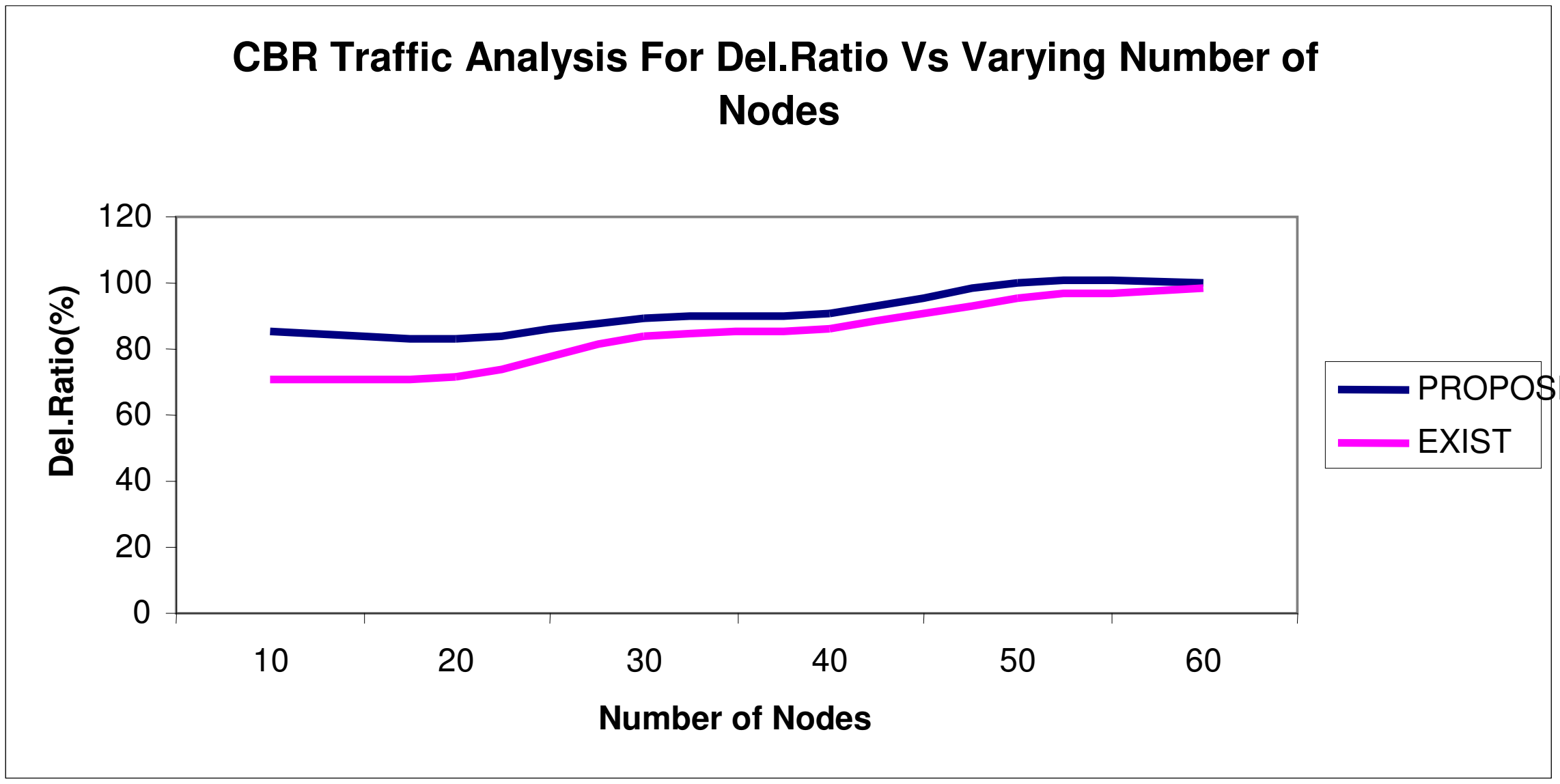

Figure XI Delivery Ratio CBR Traffic Pattern

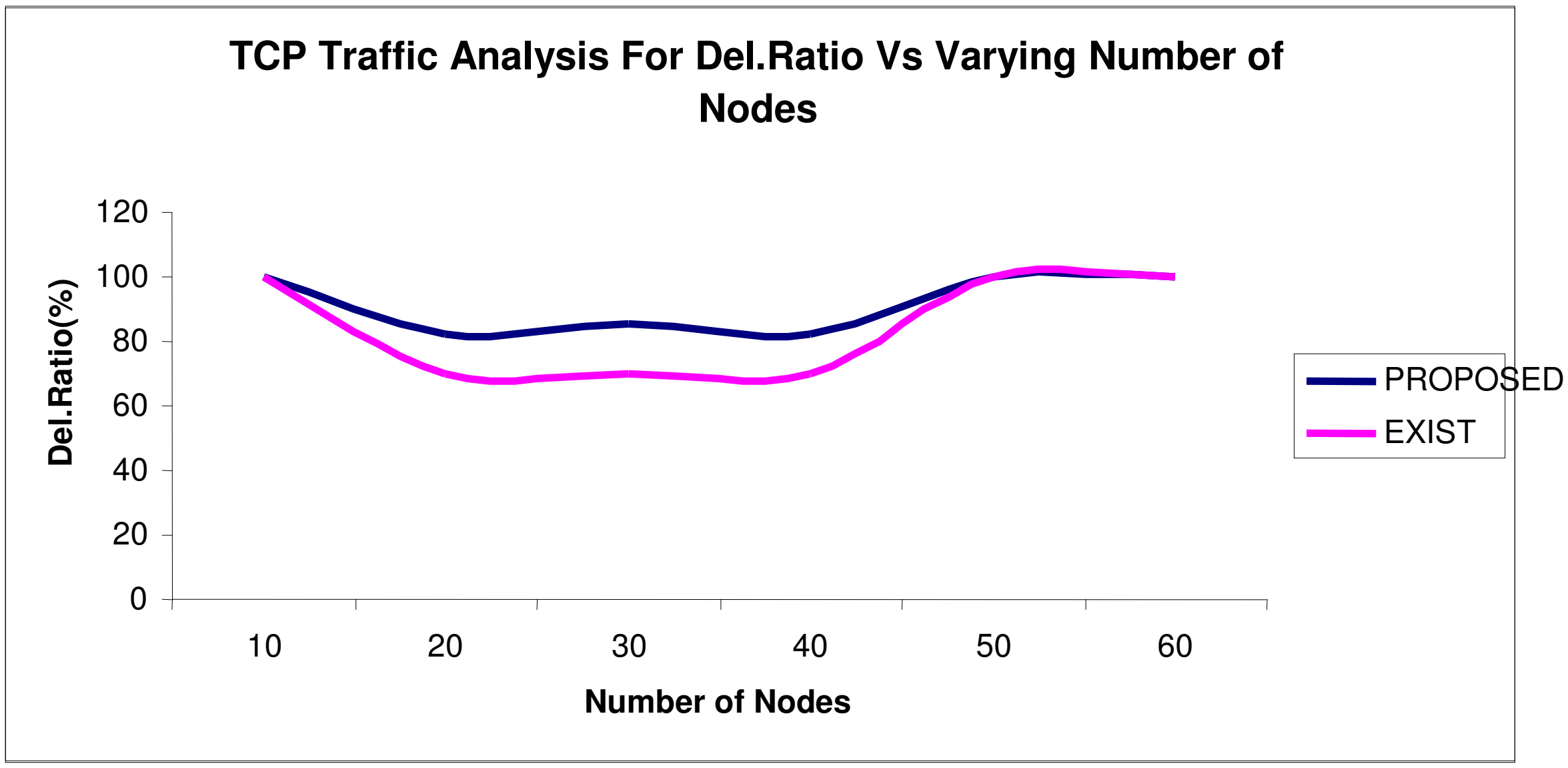

Figure XI Delivery Ratio TCP Traffic Pattern

## 7. CONCLUSION

In this paper, we have proposed a new Energy Efficient and Reliable Gossip Routing Protocol. This protocol is implemented with reduced overhead and delay and with less energy consumption. This protocol assures the increased delivery ratio, better reliability and high energy conservation for power managed routing. By simulation results, we have shown that the proposed protocol achieves good delivery ratio and good throughput with less End to End Delay and energy consumption. For further enhancement of this work, we plan to optimize the drop and queues involved in the protocol.